\documentclass[english]{article}
\usepackage[T1]{fontenc}
\usepackage[utf8]{inputenc}
\usepackage{array}
\usepackage{float}
\usepackage{amsmath}
\usepackage{graphicx}
\usepackage{authblk}

\makeatletter

\providecommand{\tabularnewline}{\\}

\makeatother

\usepackage{babel}
\begin{document}
\title{Decays $B\rightarrow D_{\left(s\right)}^{\left(*\right)}h$ ($h=\pi,\rho$)
in confined covariant quark model}

\author[1]{S. Dubni\v{c}ka}
\author[2]{A. Z. Dubni\v{c}kov\'{a}}
\author[3]{M. A. Ivanov}
\author[1]{A. Liptaj \thanks{andrej.liptaj@savba.sk}}
\affil[1]{Institute of Physics, Slovak Academy of Sciences, 845 11 Bratislava, Slovak Republic}
\affil[2]{Department of Theoretical Physics, Faculty of Mathematics, Physics and Informatics,Comenius University, 842 48 Bratislava, Slovak Republic}
\affil[3]{Bogoliubov Laboratory of Theoretical Physics, Joint Institute for Nuclear Research,141980 Dubna, Russia}

\maketitle

\begin{abstract}
Decay processes $B\rightarrow D_{\left(s\right)}^{\left(*\right)}h$ ($h=\pi,\rho$) are studied in the framework of  the confined covariant quark model using the na\"{i}ve factorization assumption. We observe that the theoretical results on branching fractions have tendency to systematically exceed the experimental numbers. Such a behavior has already been seen for similar processes by other authors.
\end{abstract}

\section{Introduction}

The recent measurements by Belle \cite{Belle:2021udv} and LHCb \cite{LHCb:2020zae}
collaborations complement the previous BaBar results \cite{BaBar:2006gzy,BaBar:2006rof,BaBar:2006zod}
on $B$ decays into $D$ particles and pions. Including also reactions
with the $\varrho$ meson in the final state \cite{LHCb:2015klp,BaBar:2006zod,CLEO:2003quc,CLEO:1994mwq},
we focus in this analysis on a set of decay processes with a rich
mix of properties. The processes differ in spin and flavor structure
and are described by various diagram topologies. Rather than addressing
some specific question, we see several broader motivations for our
study. First, we are interested in the ability of the confined covariant
quark model (CCQM) to describe the experimental branching fraction
values as established from the new measurements. The importance of
their good theoretical understanding stems from the fact that several
of the studied decay channels have a clean experimental signature
measured with a high statistical significance and thus play an important
role of a relative reference for processes, which are more difficult
to measure. Further, in the framework we use, we rely on the naïve
factorization assumption, which we in this way also indirectly test.
The assumption is presumed valid for the processes with the spectator
quark entering the $D$ meson, which is justified in the heavy-quark
limit \cite{Beneke:2000ry}. The latter can be no longer upheld if
the spectator quark becomes the part of the light meson. In addition,
our description of the chosen processes depends on five CKM matrix
elements, i.e. on all except $V_{us}$ and those related to the top
quark. So, our ability to describe various decay processes within
a single framework can be also seen as a probe and a consistency check
of the weak sector understanding. Finally, our previous works covered
most of the interesting non--leptonic $B_{\left(c,s\right)}$ decays
\cite{Dubnicka:2017job,Dubnicka:2013vm,Ivanov:2011aa,Ivanov:2002un,Ivanov:2006ni}.
$B$ decays to light unflavored mesons and $D$ particles were within
the CCQM not treated up to now.

In Sec. \ref{sec:Amplitudes-and-decay} we review the description
of the selected decays in the Standard Model a provide formulas for
amplitudes and decay widths. In Sec. \ref{sec:Hadronic-form-factors}
the main features of the CCQM are discussed and hadronic form factors
are computed. In the last section we present results and conclude
the text.

\section{Amplitudes and decay widths\label{sec:Amplitudes-and-decay}}

The work is done assuming the factorization validity and considering
the leading order Feynman diagrams.
The annihilation topologies are not taken into account, since their contributions can be neglected (see  Sec. 3.3.6 of \cite{Beneke:2000ry}).

The weak transition is described
in the effective theory approach based on a Hamiltonian constructed
from four-fermion operators $Q_{i}$ weighted by scale-dependent Wilson
coefficients $C_{i}\left(\mu\right)$ and CKM factors $V_{i}$
\[
\mathcal{H}_{\text{eff.}}=-\frac{G_{F}}{\sqrt{2}}\sum V_{i}C_{i}\left(\mu\right)Q_{i},
\]
where $G_{F}$ is the Fermi constant and $V_{i}=V_{q_{1},q_{2}}V_{q_{3},q_{4}}^{\dagger}$.
At the leading order two operators play a role
\begin{align*}
Q_{1} & =\left(\left(\overline{q_{1}}\right)_{a_{1}}\left(q_{2}\right)_{a_{1}}\right)_{V-A}\left(\left(\overline{q_{3}}\right)_{a_{2}}\left(q_{4}\right)_{a_{2}}\right)_{V-A},\\
Q_{2} & =\left(\left(\overline{q_{1}}\right)_{a_{2}}\left(q_{2}\right)_{a_{1}}\right)_{V-A}\left(\left(\overline{q_{3}}\right)_{a_{2}}\left(q_{4}\right)_{a_{1}}\right)_{V-A},\\
 & \left(\overline{q}_{1}q_{2}\right)_{V-A}\equiv\overline{q}_{1}O^{\mu}q_{2}=\overline{q}_{1}\gamma^{\mu}\left(1-\gamma^{5}\right)q_{2},
\end{align*}
where $a_{1,2}$ are color indices and $q_{i}$ represents quark fields.
Further, if $p_{B}$ denotes the four-momentum of $B$ and $p_{H}$
the momentum of the final-state meson containing the spectator quark,
then, for a given Feynman diagram, we define
\[
P=p_{B}+p_{H}\quad\text{ and }\quad q=p_{B}-p_{H}.
\]
The matrix elements of the $B\rightarrow H$ transition can be described
with help of covariant form factors where the form of the parametrization
depends on the spin of $H$. For pseudo-scalar particles ($H=PS$)
and vector particles ($H=V$) one has 
\begin{flalign*}
\left\langle PS_{\left[\bar{q}_{3},q_{2}\right]}\left(p_{H}\right)|\overline{q}_{1}O^{\mu}q_{2}|B_{\left[\bar{q}_{3},q_{1}\right]}\left(p_{b}\right)\right\rangle  & =F_{+}\left(q^{2}\right)P^{\mu}+F_{-}\left(q^{2}\right)q^{\mu},\\
\left\langle V_{\left[\bar{q}_{3},q_{2}\right]}\left(p_{H},\epsilon\right)|\overline{q}_{1}O^{\mu}q_{2}|B_{\left[\bar{q}_{3},q_{1}\right]}\left(p_{b}\right)\right\rangle  & =\frac{\epsilon_{\nu}^{\dagger}}{m_{H}+m_{V}}\\
 & \times\left[-g_{\mu\nu}P\cdot qA_{0}\left(q^{2}\right)+P^{\mu}P^{\nu}A_{+}\left(q^{2}\right)\right.\\
 & \left.+q^{\mu}P^{\nu}A_{-}\left(q^{2}\right)+i\varepsilon^{\mu\nu\alpha\beta}P_{\alpha}q_{\beta}V\left(q^{2}\right)\right].
\end{flalign*}
Because operators $Q_{1,2}$ do not contain $\sigma^{\mu\nu}q_{\nu}$
terms, the corresponding tensor form factors do enter our analysis.
It is convenient to define helicity form factors
\begin{itemize}
\item $B\rightarrow PS$
\end{itemize}
\[
H_{t}=\frac{1}{\sqrt{q^{2}}}\left[\left(m_{B}^{2}-m_{H}^{2}\right)F_{+}+q^{2}F_{-}\right],\quad H_{0}=\frac{2m_{B}\left|\mathbf{p_{2}}\right|}{\sqrt{q^{2}}}F_{+},
\]

\begin{itemize}
\item $B\rightarrow V$
\begin{align*}
H_{t} & =\frac{1}{m_{B}+m_{H}}\frac{m_{B}\left|\mathbf{p_{2}}\right|}{m_{H}\sqrt{q^{2}}}\left[\left(m_{B}^{2}-m_{H}^{2}\right)\left(A_{+}-A_{0}\right)+q^{2}A_{-}\right],\\
H_{\pm} & =\frac{1}{m_{B}+m_{H}}\left[-\left(m_{B}^{2}-m_{H}^{2}\right)A_{0}\pm2m_{B}\left|\mathbf{p_{2}}\right|V\right],\\
H_{0} & =\frac{1}{m_{B}+m_{H}}\left[-\left(m_{B}^{2}-m_{H}^{2}\right)\left(m_{B}^{2}-m_{H}^{2}-q^{2}\right)A_{0}+4m_{B}^{2}\left|\mathbf{p_{2}}\right|^{2}A_{+}\right],
\end{align*}
\end{itemize}
with $\left|\mathbf{p_{2}}\right|=\sqrt{\lambda^{\text{Källén}}\left(m_{B}^{2},m_{H}^{2},m_{H'}^{2}\right)}/2m_{B}$
being the momentum of the final state particles $H$ and $H'$ in
the rest frame of $B$. The decay width formula depends on the diagram
topology and spin structure. The studied processes can be organized
in a table with respect to the latter criteria as follows

\begin{table}[H]
\begin{centering}
\begin{tabular}{>{\centering}p{2.5cm}>{\centering}p{2.5cm}>{\centering}p{2.5cm}>{\centering}p{2.5cm}}
 & \multicolumn{3}{c}{}\tabularnewline
\hline 
\hline 
 & \multicolumn{3}{c}{Diagram type}\tabularnewline
Spin structure  & $\quad D_{1}$ & $\quad D_{2}$ & $\quad D_{3}$\tabularnewline
\hline 
 &  &  & \tabularnewline
(A)\linebreak{}
\linebreak{}
$\underline{\boldsymbol{PS}\rightarrow\boldsymbol{PS}}+PS$ & $\underline{B^{0}\rightarrow D^{-}}+\pi^{+}$\linebreak{}
$\underline{B^{0}\rightarrow\pi^{-}}+D^{+}$\linebreak{}
$\underline{B^{0}\rightarrow\pi^{-}}+D_{s}^{+}$\linebreak{}
$\underline{B^{+}\rightarrow\pi^{0}}+D_{s}^{+}$ & $\underline{B^{0}\rightarrow\pi^{0}}+\overline{D}^{0}$ & $\underline{B^{+}\rightarrow\overline{D}^{0}}+\pi^{+}$\tabularnewline
 &  &  & \tabularnewline
\hline 
 &  &  & \tabularnewline
(B)\linebreak{}
\linebreak{}
$\underline{\boldsymbol{PS}\rightarrow\boldsymbol{PS}}+V$ & $\underline{B^{0}\rightarrow D^{-}}+\varrho^{+}$\linebreak{}
$\underline{B^{0}\rightarrow\pi^{-}}+D_{s}^{*+}$\linebreak{}
$\underline{B^{+}\rightarrow\pi^{0}}+D^{*+}$\linebreak{}
$\underline{B^{+}\rightarrow\pi^{0}}+D_{s}^{*+}$ & $\underline{B^{0}\rightarrow\pi^{0}}+\overline{D^{*}}^{0}$ & $\underline{B^{+}\rightarrow\overline{D}^{0}}+\varrho^{+}$\tabularnewline
 &  &  & \tabularnewline
\hline 
 &  &  & \tabularnewline
(C)\linebreak{}
\linebreak{}
$\underline{\boldsymbol{PS}\rightarrow\boldsymbol{V}}+PS$ & $\underline{B^{0}\rightarrow D^{*-}}+\pi^{+}$\linebreak{}
$\underline{B^{0}\rightarrow\varrho^{-}}+D_{s}^{+}$\linebreak{}
$\underline{B^{+}\rightarrow\varrho^{0}}+D_{s}^{+}$ & $\underline{B^{0}\rightarrow\varrho^{0}}+\overline{D}^{0}$ & $\underline{B^{+}\rightarrow\overline{D^{*}}^{0}}+\pi^{+}$\tabularnewline
 &  &  & \tabularnewline
\hline 
 &  &  & \tabularnewline
(D)\linebreak{}
\linebreak{}
$\underline{\boldsymbol{PS}\rightarrow\boldsymbol{V}}+V$ & $\underline{B^{0}\rightarrow D^{*-}}+\varrho^{+}$

$\underline{B^{0}\rightarrow\varrho^{-}}+D_{s}^{*+}$

$\underline{B^{+}\rightarrow\varrho^{0}}+D_{s}^{*+}$ & $\underline{B^{0}\rightarrow\varrho^{0}}+\overline{D^{*}}^{0}$ & $\underline{B^{+}\rightarrow\overline{D^{*}}^{0}}+\varrho^{+}$\tabularnewline
 &  &  & \tabularnewline
\hline 
\hline 
 &  &  & \tabularnewline
\end{tabular}
\par\end{centering}
\caption{Classification of processes.}
\label{classification}
\end{table}

\noindent \begin{flushleft}
where the diagram types are depicted in the following Figure
\par\end{flushleft}

\begin{flushleft}
\begin{figure}[H]
\begin{centering}
\includegraphics[width=0.4\textwidth]{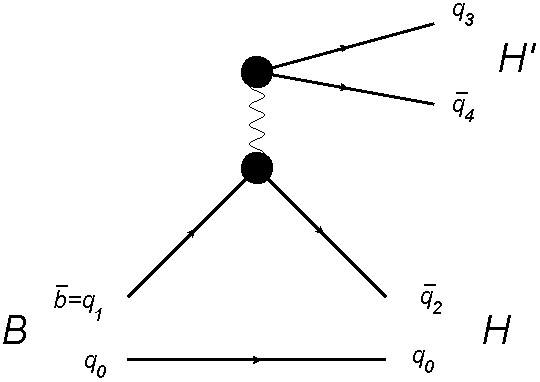}\hspace{1cm}\includegraphics[width=0.4\textwidth]{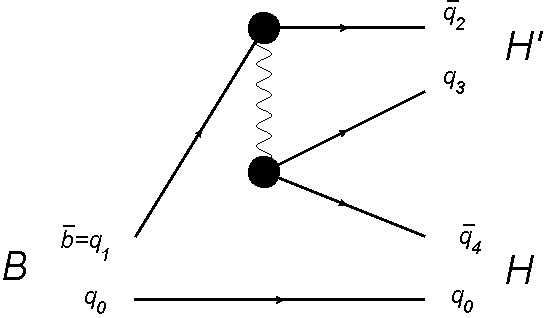}
\par\end{centering}
\begin{centering}
($D_{1}$)\hspace{6cm}($D_{2}$)\vspace{1cm}
\par\end{centering}
\begin{centering}
\includegraphics[width=0.9\textwidth]{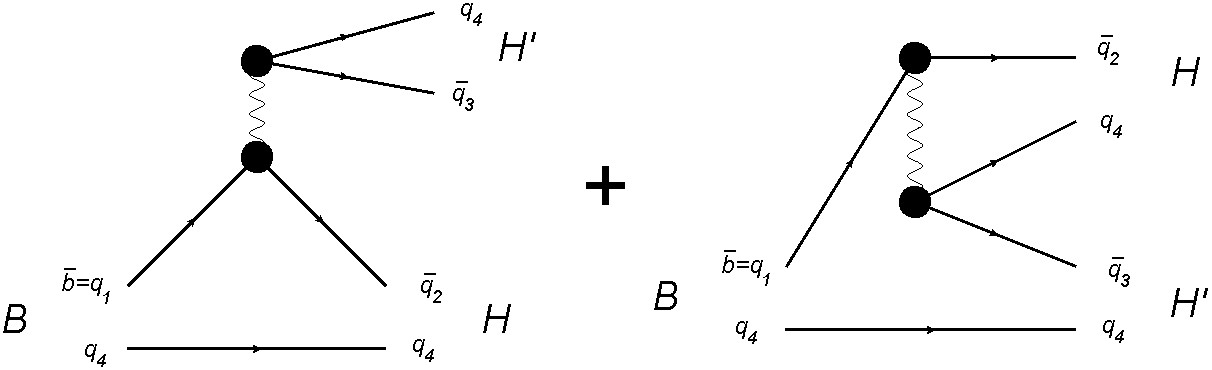}
\par\end{centering}
\begin{centering}
($D_{3}$)
\par\end{centering}
\caption{Diagram structures $D_{1}$, $D_{2}$ and $D_{3}$.}

\end{figure}
\par\end{flushleft}

\noindent For the processes listed in Table \ref{classification}
we underline the part containing the transition of the spectator quark,
for the $D_{3}$ case we apply this rule the fist of the two diagrams.
The $D_{1/2/3}$ decays are sometimes referred to as class-1/2/3 processes
\cite{Neubert:2001sj}, $D_{2}$ are called color-suppressed. In our
analysis we do not distinguish between $u$ and $d$ quarks and denote
both by $q$. In order to make the decay width formulas compact we
introduce the symbol $\theta_{0}$ which takes the value $1/\sqrt{2}$
if an unflavored light neutral meson is in the final state and is
equal to one otherwise. The formulas for the $D_{1}$ decays are
\[
\Gamma_{[A,D_{1}]}\left(\underline{S_{1}\rightarrow S_{2}}+S_{3}\right)=\frac{G_{F}^{2}}{16\pi}\frac{\left|\boldsymbol{p}_{2}\right|}{m_{S_{1}}^{2}}\left|\theta_{0}V_{q_{1}q_{2}}V_{q_{3}q_{4}}^{\dagger}a_{1}f_{S_{3}}m_{S_{3}}\right|^{2}\left\{ H_{t}^{S_{1}\rightarrow S_{2}}\left(m_{S_{3}}^{2}\right)\right\} ^{2},
\]
\[
\Gamma_{[B,D_{1}]}\left(\underline{S_{1}\rightarrow S_{2}}+V\right)=\frac{G_{F}^{2}}{16\pi}\frac{\left|\boldsymbol{p}_{2}\right|}{m_{S_{1}}^{2}}\left|\theta_{0}V_{q_{1}q_{2}}V_{q_{3}q_{4}}^{\dagger}a_{1}f_{V}m_{V}\right|^{2}\left\{ H_{0}^{S_{1}\rightarrow S_{2}}\left(m_{V}^{2}\right)\right\} ^{2},
\]
\[
\Gamma_{[C,D_{1}]}\left(\underline{S_{1}\rightarrow V}+S_{2}\right)=\frac{G_{F}^{2}}{16\pi}\frac{\left|\boldsymbol{p}_{2}\right|}{m_{S_{1}}^{2}}\left|\theta_{0}V_{q_{1}q_{2}}V_{q_{3}q_{4}}^{\dagger}a_{1}f_{S_{2}}m_{S_{2}}\right|^{2}\left\{ H_{t}^{S_{1}\rightarrow V}\left(m_{S_{2}}^{2}\right)\right\} ^{2},
\]
\[
\Gamma_{[D,D_{1}]}\left(\underline{S\rightarrow V_{1}}+V_{2}\right)=\frac{G_{F}^{2}}{16\pi}\frac{\left|\boldsymbol{p}_{2}\right|}{m_{S}^{2}}\left|\theta_{0}V_{q_{1}q_{2}}V_{q_{3}q_{4}}^{\dagger}a_{1}f_{V_{2}}m_{V_{2}}\right|^{2}\sum_{i=0,\pm}\left\{ H_{i}^{S_{1}\rightarrow V_{1}}\left(m_{V_{2}}^{2}\right)\right\} ^{2}.
\]
The decay formulas for the color-suppressed $D_{2}$ processes can
be written as function of those for $D_{1}$
\[
\Gamma_{[X,D_{2}]}^{[a_{1},a_{2}]}\left(\underline{B\rightarrow H}+H'\right)=\Gamma_{[X,D_{1}]}^{[a_{2},a_{1}]}\left(\underline{B\rightarrow H}+H'\right),
\]
where we make appear the coefficients $a_{1}$ and $a_{2}$ which
we define hereunder and whose roles are swapped. The decay widths
of the $D_{3}$ processes are given by
\begin{align*}
\Gamma_{[A,D_{3}]}\left(\underline{S_{1}\rightarrow S_{2}}+S_{3}\right) & =\frac{G_{F}^{2}}{16\pi}\frac{\left|\boldsymbol{p}_{2}\right|}{m_{S_{1}}^{2}}\theta_{0}^{2}\left|V_{q_{1}q_{2}}V_{q_{3}q_{4}}^{\dagger}\right|^{2}\\
 & \times\left\{ a_{1}f_{S_{3}}m_{S_{3}}H_{t}^{S_{1}\rightarrow S_{2}}\left(m_{S_{3}}^{2}\right)+a_{2}f_{S_{2}}m_{S_{2}}H_{t}^{S_{1}\rightarrow S_{3}}\left(m_{S_{2}}^{2}\right)\right\} ^{2},
\end{align*}
\begin{align*}
\Gamma_{[B,D_{3}]}\left(\underline{S_{1}\rightarrow S_{2}}+V\right) & =\frac{G_{F}^{2}}{16\pi}\frac{\left|\boldsymbol{p}_{2}\right|}{m_{S_{1}}^{2}}\theta_{0}^{2}\left|V_{q_{1}q_{2}}V_{q_{3}q_{4}}^{\dagger}\right|^{2}\\
 & \times\left\{ a_{1}f_{V}m_{V}H_{0}^{S_{1}\rightarrow S_{2}}\left(m_{V}^{2}\right)+a_{2}f_{S_{2}}m_{S_{2}}H_{t}^{S_{1}\rightarrow V}\left(m_{S_{2}}^{2}\right)\right\} ^{2}
\end{align*}
\begin{align*}
\Gamma_{[C,D_{3}]}\left(\underline{S_{1}\rightarrow V}+S_{2}\right) & =\frac{G_{F}^{2}}{16\pi}\frac{\left|\boldsymbol{p}_{2}\right|}{m_{S_{1}}^{2}}\theta_{0}^{2}\left|V_{q_{1}q_{2}}V_{q_{3}q_{4}}^{\dagger}\right|^{2}\\
 & \times\left\{ a_{1}f_{S_{2}}m_{S_{2}}H_{t}^{S_{1}\rightarrow V}\left(m_{S_{2}}^{2}\right)+a_{2}f_{V}m_{V}H_{0}^{S_{1}\rightarrow S_{2}}\left(m_{V}^{2}\right)\right\} ^{2},
\end{align*}
\begin{align*}
\Gamma_{[D,D_{3}]}\left(\underline{S\rightarrow V_{1}}+V_{2}\right) & =\frac{G_{F}^{2}}{16\pi}\frac{\left|\boldsymbol{p}_{2}\right|}{m_{S}^{2}}\theta_{0}^{2}\left|V_{q_{1}q_{2}}V_{q_{3}q_{4}}^{\dagger}\right|^{2}\\
 & \times\sum_{i=0,\pm}\left\{ a_{1}f_{V_{2}}m_{V_{2}}H_{i}^{S\rightarrow V_{1}}\left(m_{V_{2}}^{2}\right)+a_{2}f_{V_{1}}m_{V_{1}}H_{i}^{S\rightarrow V_{2}}\left(m_{V_{1}}^{2}\right)\right\} ^{2}.
\end{align*}
In the above formulas $f$ denotes the leptonic decay constant and
$a_{1}$ and $a_{2}$ are combinations of the Wilson coefficients
$C_{1}$ and $C_{2}$

\noindent 
\[
a_{1}=C_{1}+\xi C_{2},\quad a_{2}=C_{2}+\xi C_{1}.
\]
Here $\xi$ is the color suppression factor inversely proportional
to the number of colors $N_{c}$, $\xi=1/N_{c}$. Working in the large-$N_{c}$
limit $\xi=0$ we have
\[
a_{1}=1.0111\quad a_{2}=-0.2632.
\]
We take values of the Wilson coefficients from \cite{Descotes-Genon:2013vna},
where they were computed at the matching scale $\mu_{0}=2M_{W}$ at
the NNLO precision and run down to the

\noindent hadronic scale $\mu_{b}=4.8\mathrm{GeV}$. The last ingredient
necessary for the evaluation of decay widths are the hadronic form
factors. Because of their non-perturbative nature, one has to rely
on a model-dependent approach. We evaluate these form factors within
the CCQM.

\section{Hadronic form factors in CCQM\label{sec:Hadronic-form-factors}}

The description of nonleptonic heavy meson decays in the framework
of the CCQM was already presented several times \cite{Dubnicka:2017job,Dubnicka:2013vm,Ivanov:2011aa,Ivanov:2002un,Ivanov:2006ni,ATLAS:2022aiy}.
We summarize here the most important attributes of our approach. 

The CCQM uses the scheme where a hadron is before the interaction
converted into its constituent quarks. This is expressed by the non-local
effective Lagrangian
\begin{align*}
\mathcal{L}_{int} & =g_{M}M\left(x\right)J_{M}\left(x\right)+\text{ H.c. },\\
J_{M}\left(x\right) & =\int dx_{1}\int dx_{2}F_{M}\left(x;x_{1},x_{2}\right)\overline{q}_{2}\left(x_{2}\right)\varGamma_{M}q_{1}\left(x_{1}\right),
\end{align*}
which guarantees a full Lorentz covariance. The interaction strength
between the mesonic field $M$ and its interpolating quark current
$J_{M}$ is given by the coupling $g_{M}$. The current is constructed
from quark fields $q$, an appropriate Dirac matrix $\varGamma_{M}$
and a vertex function $F_{M}$. The latter is chosen to have a translational
invariant form
\[
F_{M}\left(x;x_{1},x_{2}\right)=\delta\left(x-w_{1}x_{1}-w_{2}x_{2}\right)\varPhi_{M}\left[\left(x_{1}-x_{2}\right)^{2}\right]
\]
with $w_{i}=m_{q_{i}}/(m_{q_{1}}+m_{q_{2}})$, so that the meson position
$x$ can be interpreted as the barycenter of the quark system. The
function $\varPhi_{M}$ is taken Gaussian in the momentum representation
\[
\varPhi_{M}\left[\left(x_{1}-x_{2}\right)^{2}\right]=\int\frac{d^{4}k}{\left(2\pi\right)^{4}}e^{-ik\left(x_{1}-x_{2}\right)}\widetilde{\varPhi}_{M}\left(-k^{2}\right),\quad\widetilde{\varPhi}_{M}\left(-k^{2}\right)=e^{k^{2}/\varLambda_{M}^{2}}.
\]
Here $\varLambda_{M}$ is a free parameter of the model which characterizes
the meson $M$.

The presence of both, hadrons and quarks, rises concerns about the
double counting. We remedy the latter by applying the so-called compositeness
condition \cite{Ganbold:2014pua}
\begin{equation}
Z_{M}=1-g_{M}^{2}\Pi_{M}^{'}\left(m_{M}^{2}\right)=0,\label{eq:Compositeness}
\end{equation}
which originates in the works \cite{Jouvet:1956ii,Salam:1962ap,Weinberg:1962hj}.
Here $\Pi_{M}^{'}$ is the derivative of the mass operator corresponding
to the self--energy diagram of the meson field fluctuating into a
pair of quarks. Setting the renormalization constant $Z_{M}^{1/2}$
to zero implies that the physical and the corresponding bare state
have no overlap, i.e. the physical state does not contain the bare
state and is therefore interpreted as bound. The condition effectively
excludes the constituent degrees of freedom from the space of physical
states because the constituents exist in virtual states only. The
equality in (\ref{eq:Compositeness}) is reached by an appropriate
choice of $g_{M}$ and in this way the coupling constants are determined
and do not appear as model parameters.

Another notable feature of the CCQM is the confining property. So
as to prevent hadrons from decaying into quarks in situations where
the hadron mass is greater than those of constituent quarks summed,
an infrared cutoff $1/\lambda^{2}$ is introduced as an upper integration
limit in the integration over the space of Schwinger parameters. The
latter appear in the parameterization of quark propagators, which
become, after the cutoff being applied, entire functions with all
possible thresholds in the corresponding quark loop diagrams removed
(more details given in \cite{Branz:2009cd}, Section II C).

\begin{figure}[t]
\begin{centering}
\includegraphics[width=0.47\textwidth]{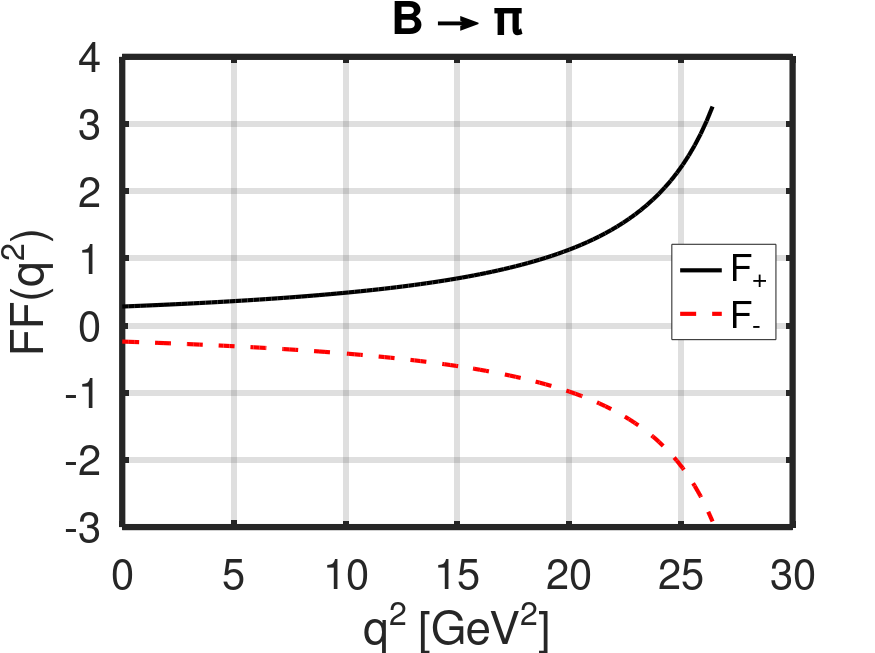}\qquad{}\includegraphics[width=0.47\textwidth]{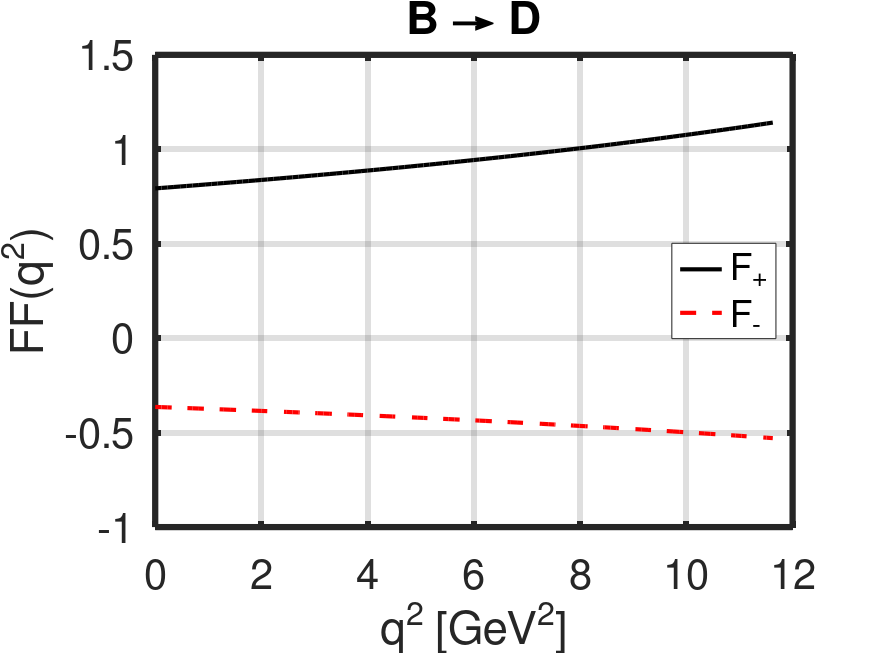}\bigskip{}
\par\end{centering}
\centering{}\includegraphics[width=0.47\textwidth]{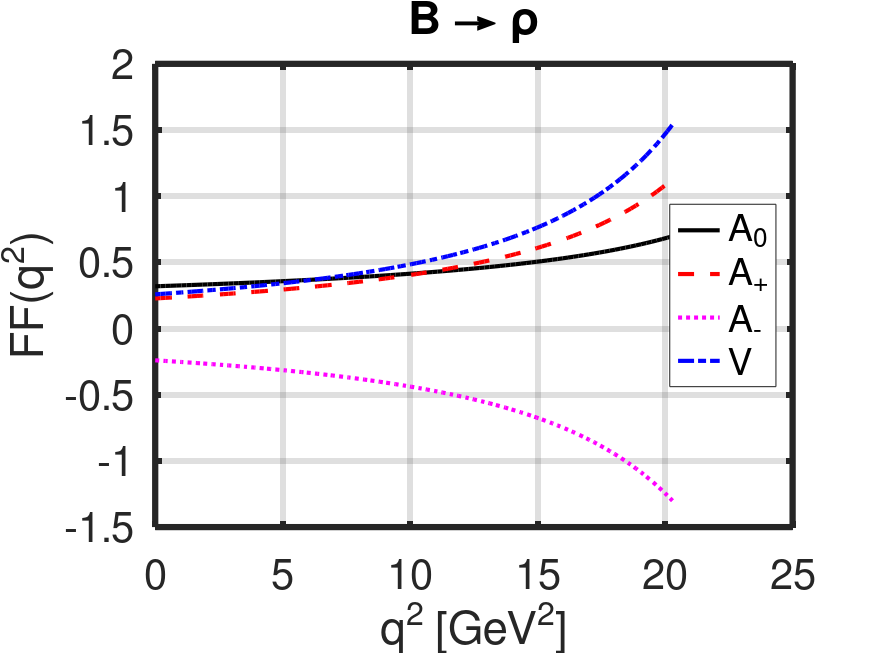}\qquad{}\includegraphics[width=0.47\textwidth]{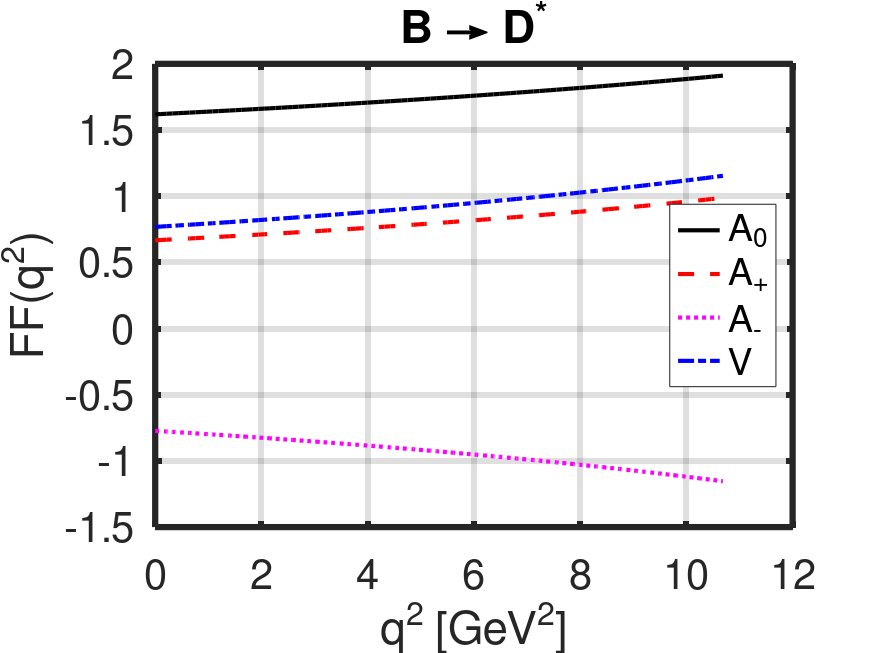}\caption{Form factors of transitions which determine the investigated decays.}
\label{fig:FFs}
\end{figure}
The evaluation of hadronic form factors within the CCQM proceeds via
standard computation techniques based on evaluation of the corresponding
Feynman diagrams. For $B\rightarrow PS$ and $B\rightarrow V$ the
transition the form factors are given by
\begin{align*}
 & F_{+}\left(q^{2}\right)P^{\mu}+F_{-}\left(q^{2}\right)q^{\mu}=\\
 & \hspace{2cm}=\frac{3g_{B}g_{H}}{4\pi^{2}}\int\frac{d^{4}k}{4\pi^{2}i}\widetilde{\varPhi}_{B}\left[-\left(k+w_{b}p_{B}\right)^{2}\right]\widetilde{\varPhi}_{H}\left[-\left(k+w_{q_{i}}p_{H}\right)^{2}\right]\\
 & \hspace{2.5cm}\times\mathrm{Tr}\left[S_{q_{i}}\left(k+p_{H}\right)O^{\mu}S_{b}\left(k+p_{B}\right)\gamma^{5}S_{q_{s}}\left(k\right)\gamma^{5}\right],
\end{align*}
\begin{align*}
 & \frac{\epsilon_{H\nu}^{*}}{m_{B}+m_{H}}[-g^{\mu\nu}PqA_{0}\left(q^{2}\right)+P^{\mu}P^{\nu}A_{+}\left(q^{2}\right)\\
 & +q^{\mu}P^{\nu}A_{-}\left(q^{2}\right)+i\varepsilon^{\mu\nu\alpha\beta}P_{\alpha}q_{\beta}V\left(q^{2}\right)]=\\
 & \hspace{2cm}=\frac{3g_{B}g_{H}}{4\pi^{2}}\int\frac{d^{4}k}{4\pi^{2}i}\widetilde{\varPhi}_{B}\left[-\left(k+w_{b}p_{B}\right)^{2}\right]\widetilde{\varPhi}_{H}\left[-\left(k+w_{q_{i}}p_{H}\right)^{2}\right]\\
 & \hspace{2.5cm}\times\mathrm{Tr}\left[S_{q_{i}}\left(k+p_{H}\right)O^{\mu}S_{b}\left(k+p_{B}\right)\gamma^{5}S_{q_{s}}\left(k\right)\epsilon_{H\mu}\gamma^{\mu}\right].
\end{align*}
Here $q_{s}$ and $q_{i}$ denote the spectator and the interacting
quark of $H$ respectively ($H$ being $PS$ or $V$), $S$ represents
quark propagators and $\epsilon_{H}$ the polarization vector of $H\equiv V$.
Giving the vertex functions the above-mentioned Gaussian form and
writing the propagators in the Schwinger representation, one performs
the loop integration and applies the infrared cutoff in the integration
over the Schwinger parameters, this last integration being done numerically.
The model parameters were determined in our previous works \cite{Ivanov:2015tru,Ganbold:2014pua}
and their numerical values are
\begin{table}[H]
\begin{centering}
\begin{tabular}{ccccccccccc}
\hline 
$\varLambda_{B}$ & $\varLambda_{D}$ & $\varLambda_{D^{*}}$ & $\varLambda_{D_{s}}$ & $\varLambda_{D_{s}^{*}}$ & $\varLambda_{\pi}$ & $\varLambda_{\rho}$ & $m_{b}$ & $m_{s}$ & $m_{q}$ & $\lambda$\tabularnewline
$1.96$ & $1.60$ & $1.53$ & $1.75$ & $1.56$ & $0.87$ & $0.61$ & $5.04$ & $0.428$ & $0.241$ & $0.181$\tabularnewline
\hline 
\end{tabular}\caption{CCQM parameters. Values are in GeV.}
\par\end{centering}
\end{table}
 The predicted behavior of form factors in four studied transitions
is shown in Fig. \ref{fig:FFs}.

\section{Results, conclusion}

\begin{table}[t]
\begin{centering}
\begin{tabular}{ccccccc}
 &  &  &  &  &  & \tabularnewline
\hline 
\hline 
 & Process & Diagram & $\mathcal{B}_{\mathrm{CCQM}}/\mathrm{E}$ &  & $\mathcal{B}_{\mathrm{PDG}}/\mathrm{E}$ & $\mathrm{E}$\tabularnewline
\hline 
1 & $B^{0}\rightarrow D^{-}+\pi^{+}$ & $D_{1}$ & $5.34 \pm 0.27$ &  & $2.52\pm0.13$ & $10^{-3}$\tabularnewline
2 & $B^{0}\rightarrow\pi^{-}+D^{+}$ & $D_{1}$ & $11.19 \pm 0.56$ &  & $7.4\pm1.3$ & $10^{-7}$\tabularnewline
3 & $B^{0}\rightarrow\pi^{-}+D_{s}^{+}$ & $D_{1}$ & $3.48 \pm 0.17$ &  & $2.16\pm0.26$ & $10^{-5}$\tabularnewline
4 & $B^{+}\rightarrow\pi^{0}+D_{s}^{+}$ & $D_{1}$ & $1.88 \pm 0.09$ &  & $1.6\pm0.5$ & $10^{-5}$\tabularnewline
5 & $B^{0}\rightarrow D^{-}+\rho^{+}$ & $D_{1}$ & $14.06 \pm 0.70$ &  & $7.6\pm1.2$ & $10^{-3}$\tabularnewline
6 & $B^{0}\rightarrow\pi^{-}+D_{s}^{*+}$ & $D_{1}$ & $3.66 \pm 0.18$ &  & $2.1\pm0.4$ & $10^{-5}$\tabularnewline
7 & $B^{+}\rightarrow\pi^{0}+D^{*+}$ & $D_{1}$ & $0.804 \pm 0.04$ &  & $<3.6$ & $10^{-6}$\tabularnewline
8 & $B^{+}\rightarrow\pi^{0}+D_{s}^{*+}$ & $D_{1}$ & $0.197 \pm 0.01$ &  & $<2.6$ & $10^{-4}$\tabularnewline
9 & $B^{0}\rightarrow D^{*-}+\pi^{+}$ & $D_{1}$ & $4.74 \pm 0.24$ &  & $2.74\pm0.13$ & $10^{-3}$\tabularnewline
10 & $B^{0}\rightarrow\rho^{-}+D_{s}^{+}$ & $D_{1}$ & $2.76 \pm 0.14$ &  & $<2.4$ & $10^{-5}$\tabularnewline
11 & $B^{+}\rightarrow\rho^{0}+D_{s}^{+}$ & $D_{1}$ & $0.149 \pm 0.01$ &  & $<3.0$ & $10^{-4}$\tabularnewline
12 & $B^{0}\rightarrow D^{*-}+\rho^{+}$ & $D_{1}$ & $14.58 \pm 0.73$ &  & $6.8\pm0.9$ & $10^{-3}$\tabularnewline
13 & $B^{0}\rightarrow\rho^{-}+D_{s}^{*+}$ & $D_{1}$ & $5.09 \pm 0.25$ &  & $4.1\pm1.3$ & $10^{-5}$\tabularnewline
14 & $B^{+}\rightarrow\rho^{0}+D_{s}^{*+}$ & $D_{1}$ & $0.275 \pm 0.01$ &  & $<4.0$ & $10^{-4}$\tabularnewline
\cline{2-7} \cline{3-7} \cline{4-7} \cline{5-7} \cline{6-7} \cline{7-7} 
15 & $B^{0}\rightarrow\pi^{0}+\overline{D}^{0}$ & $D_{2}$ & $0.085 \pm 0.00$ &  & $2.63\pm0.14$ & $10^{-4}$\tabularnewline
16 & $B^{0}\rightarrow\pi^{0}+\overline{D}^{*0}$ & $D_{2}$ & $1.13 \pm 0.06$ &  & $2.2\pm0.6$ & $10^{-4}$\tabularnewline
17 & $B^{0}\rightarrow\rho^{0}+\overline{D}^{0}$ & $D_{2}$ & $0.675 \pm 0.03$ &  & $3.21\pm0.21$ & $10^{-4}$\tabularnewline
18 & $B^{0}\rightarrow\rho^{0}+\overline{D}^{*0}$ & $D_{2}$ & $1.50 \pm 0.08$ &  & $<5.1$ & $10^{-4}$\tabularnewline
\cline{2-7} \cline{3-7} \cline{4-7} \cline{5-7} \cline{6-7} \cline{7-7} 
19 & $B^{+}\rightarrow\overline{D}^{0}+\pi^{+}$ & $D_{3}$ & $3.89 \pm 0.19$ &  & $4.68\pm0.13$ & $10^{-3}$\tabularnewline
20 & $B^{+}\rightarrow\overline{D}^{0}+\rho^{+}$ & $D_{3}$ & $1.83 \pm 0.09$ &  & $1.34\pm0.18$ & $10^{-2}$\tabularnewline
21 & $B^{+}\rightarrow\overline{D}^{*0}+\pi^{+}$ & $D_{3}$ & $7.60 \pm 0.38$ &  & $4.9\pm0.17$ & $10^{-3}$\tabularnewline
22 & $B^{+}\rightarrow\overline{D}^{*0}+\rho^{+}$ & $D_{3}$ & $11.75 \pm 0.59$ &  & $9.8\pm1.7$ & $10^{-3}$\tabularnewline
\hline 
\hline 
 &  &  &  &  &  & \tabularnewline
\end{tabular}
\par\end{centering}
\caption{Computed branching fractions compared with experimental measurements
\cite{ParticleDataGroup:2020ssz}.}
\label{Tab:results}
\end{table}
\begin{figure}
\begin{centering}
\includegraphics[width=0.75\linewidth]{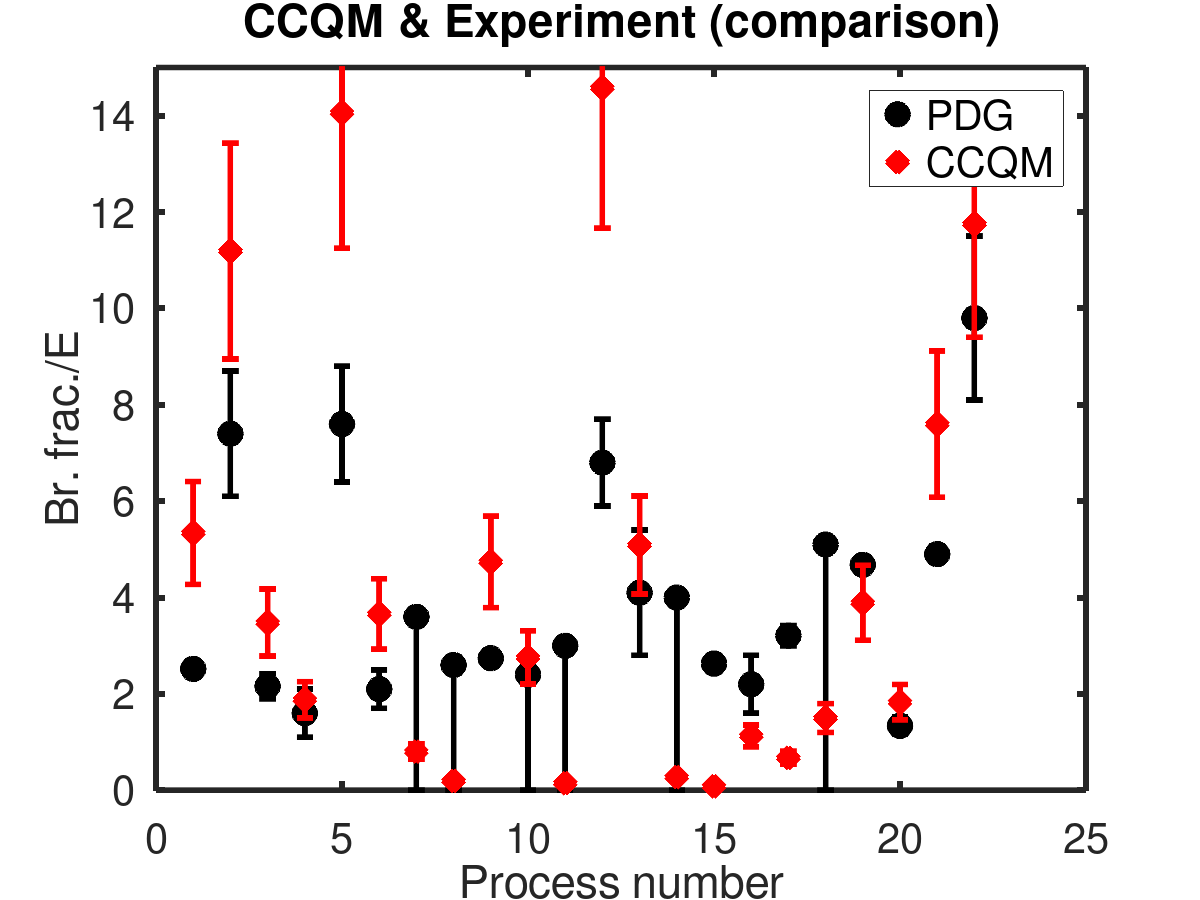}
\par\end{centering}
\caption{CCQM branching fraction predictions compared with the experimental
values listed in the PDG \cite{ParticleDataGroup:2020ssz}. Processes
are numbered as in Tab. \ref{Tab:results}.}
\label{fig:ccqmPDG}
\end{figure}
Our results are summarized in Table \ref{Tab:results}, the error on branching fractions is estimated to be 20\%. The precision of the description of the experimental data is limited, as seen in
Fig \ref{fig:ccqmPDG}. The central values of the CCQM numbers are in agreement with those
measurements which provide upper limits and in two other cases they
lay inside 1$\sigma$ error of the measured value. For the rest, the
CCQM provides mostly fair estimates of the experimental numbers, usually
within the factor of two. However, also in these situations the difference
in terms of standard deviations can be large, if the measured point
has a small error. From this point of view large deviations are seen
for color-suppressed processes\footnote{We use process numbers as in Table \ref{Tab:results}.}
(15)(17) and for those with $B\rightarrow D^{\left(*\right)}$ transition
(1)(9)(21). Since the factorization assumption has no solid justification
for the color-suppressed decays we address, we attribute the observe
difference in (15)(17) to its breaking. Concerning processes (1)(9)(21),
they have rather small experimental errors which may partly explain
the large differences in terms of sigmas. One may also notice that
they share the same set of form factors which implies correlation
in their behavior and one indeed observes important overestimation
also for other $B\rightarrow D^{\left(*\right)}$ processes, e.g.
(5)(12). Actually, an overestimation is seen for almost all $D_{1}$
and $D_{3}$ decays (with the exception of (19)), the overestimation
is just more pronounced for some processes than for others. Such systematic
shift is somewhat surprising, but we are not the first to observe
it, see \cite{Huber:2016xod, Bordone:2020gao}. The authors of \cite{Iguro:2020ndk}
 notice the same behavior in similar decays too. They argue that it
is difficult to provide a solid explanation within the Standard Model
and thus propose new physics mechanisms. Our results seem to confirm their
observations, which they label as ``novel puzzle''. New physics explanations are also investigated in \cite{Cai:2021mlt}. The comparison of our results with those of other authors is
shown in Table \ref{tab:comparisonWithOthers}.
\begin{table}
\begin{centering}
\begin{tabular}{cccc}
\hline 
\hline 
Decay mode & our results & \cite{Huber:2016xod} & \cite{Cai:2021mlt} \tabularnewline
\hline 
$B^{0}\rightarrow D^{-}\pi^{+}$ & $5.34\pm0.27$ & $3.93_{-0.42}^{+0.43}$ & $4.74_{-0.69}^{+0.61}$\tabularnewline
$B^{0}\rightarrow D^{*-}\pi^{+}$ & $4.74\pm0.24$ & $3.45_{-0.50}^{+0.53}$ & $4.26_{-0.80}^{+0.75}$\tabularnewline
$B^{0}\rightarrow D^{-}\rho^{+}$ & $14.06\pm0.70$ & $10.42_{-1.20}^{+1.24}$ & $12.28_{-1.63}^{+1.40}$\tabularnewline
$B^{0}\rightarrow D^{*-}\rho^{+}$ & $14.58\pm0.73$ & $9.24_{-0.71}^{+0.72}$ & $11.61_{-2.01}^{+1.88}$\tabularnewline
\hline 
\hline 
\end{tabular}
\par\end{centering}
\caption{Comparison of theoretical predictions for chosen branching fractions
(in units $10^{-3}$).\label{tab:comparisonWithOthers}}

\end{table}

\section*{Acknowledgement}

S. D. , A. Z. D. and A. L. acknowledge the support from the Scientific
Grant Agency VEGA, Grant No. 2/0105/21. All authors acknowledge the
support from the Joint Research Project of the Institute of Physics,
Slovak Academy of Sciences and the Bogoliubov Laboratory of Theoretical
Physics, Joint Institute for Nuclear Research, Grant No. 01-3-1135-2019/2023.

\bibliographystyle{unsrt}
\bibliography{text_BtoDh_v1}

\end{document}